\documentclass[reprint,amsmath,amssymb,aps]{revtex4-1}

\usepackage{graphicx}
\usepackage{dcolumn}
\usepackage{bm}
\usepackage{hyperref}
\usepackage{epstopdf}
\usepackage{eucal}

\usepackage[sort&compress]{natbib}
\graphicspath{{Figures/}{./Figures/}}

\begin{document}

\title{Deep reinforcement learning for optical systems:  A case study of mode-locked lasers}

\author{Chang Sun$^*$, Eurika Kaiser$^{**}$, Steven L. Brunton$^{**,\dag}$, and J. Nathan Kutz$^{*,\dag}$\\[.2in]
{\em $^*$ Department of Physics, University of Washington, Seattle, WA 98115}\\
{\em $^{**}$ Department of Mechanical Engineering, University of Washington, Seattle, WA 98115}\\
{\em $^\dag$ Department of Applied Mathematics, University of Washington, Seattle, WA 98115} }

\begin{abstract}
We demonstrate that deep reinforcement learning (deep RL) provides a highly effective strategy for the control and self-tuning of optical systems.  Deep RL integrates the two leading machine learning architectures of deep neural networks and reinforcement learning to produce robust and stable learning for control.  Deep RL is ideally suited for optical systems as the tuning and control relies on interactions with its environment with a goal-oriented objective  to achieve optimal immediate or delayed rewards.  This allows the optical system to recognize bi-stable structures and navigate, via trajectory planning, to optimally performing solutions, the first such algorithm demonstrated to do so in optical systems.  We specifically demonstrate the deep RL architecture on a mode-locked laser, where robust self-tuning and control can be established through access of the deep RL agent to its waveplates and polarizers.  We further integrate transfer learning to help the deep RL agent rapidly learn new parameter regimes and generalize its control authority.  Additionally, the deep RL learning can be easily integrated with other control paradigms to provide a broad framework to control any optical system.
\end{abstract}

 \maketitle


Machine learning (ML) and artificial intelligence (AI) algorithms are transforming the scientific landscape~\cite{goodfellow2016deep,brunton2019data}.  From self-driving cars and autonomous vehicles to digital twins and manufacturing, there are few scientific and engineering disciplines that have not been profoundly impacted by the rise of ML/AI methods.   Optics is no exception, with a significant growth of ML/AI methods developed for applications ranging from imaging to optical communications~\cite{zibar2017machine,won2018intelligent}.   For control applications,  a variety of ML strategies have been developed for stabilizing optical systems such as mode-locked lasers~\cite{brunton2013extremum,brunton2014self,fu2014classification,kutz2015intelligent,andral2015fiber,woodward2016towards,andral2016toward,baumeister2018deep}.   From genetic algorithms to deep neural networks, these studies provide a broad perspective on how a diverse set of optimization algorithms can be used to automate the control and self-tuning of a given optical device.   However, one of the most successful ML architectures has yet to be implemented for mode-locked lasers:  {\em reinforcement learning} (RL)~\cite{sutton2018reinforcement}.  RL is a rapidly growing branch of ML/AI that is based upon goal-oriented algorithms in which an agent learns from interactions with the environment.  It is the algorithmic basis for the popular AI success stories on games like chess and Go~\cite{silver2018general}.  Given its leading status as a control and goal-oriented strategy, we show that RL can be integrated with optical systems, specifically mode-locked lasers, to produce an architecture for intelligent and stable self-tuning operation.

The power of RL lies in its ability to learn from interactions with the environment with goal-oriented objectives.  This is unlike the two other dominant ML paradigms of supervised and unsupervised learning~\cite{goodfellow2016deep,brunton2019data}.  With a trial-and-error search, a RL agent learns to sense the state of its environment and take actions accordingly to achieve optimal immediate or delayed rewards. Specifically, the RL agent arrives at different states by performing actions, with the selected actions leading to positive or negative rewards for learning.  Importantly, the RL agent is capable of learning delayed rewards, which is critical for many optical systems since a trajectory to the optimal solution must be learned.  This is equivalent to mapping out a set of moves, or long term strategy, to win a chess game.  RL  targets optimal policies for reinforcement learners to maximize the total reward across an episode.   Each state follows a Markov property by assumption, i.e., each state is determined only by the previous state and the transition taken to the current state.   Thus a large number of trials must be evaluated in order to determine an optimal policy.  This is accomplished in chess and Go by self-play~\cite{silver2018general}, which is exactly what the mode-locked laser is allowed to do to learn an optimal policy.

\begin{figure}[t]
 \includegraphics[width=0.45\textwidth]{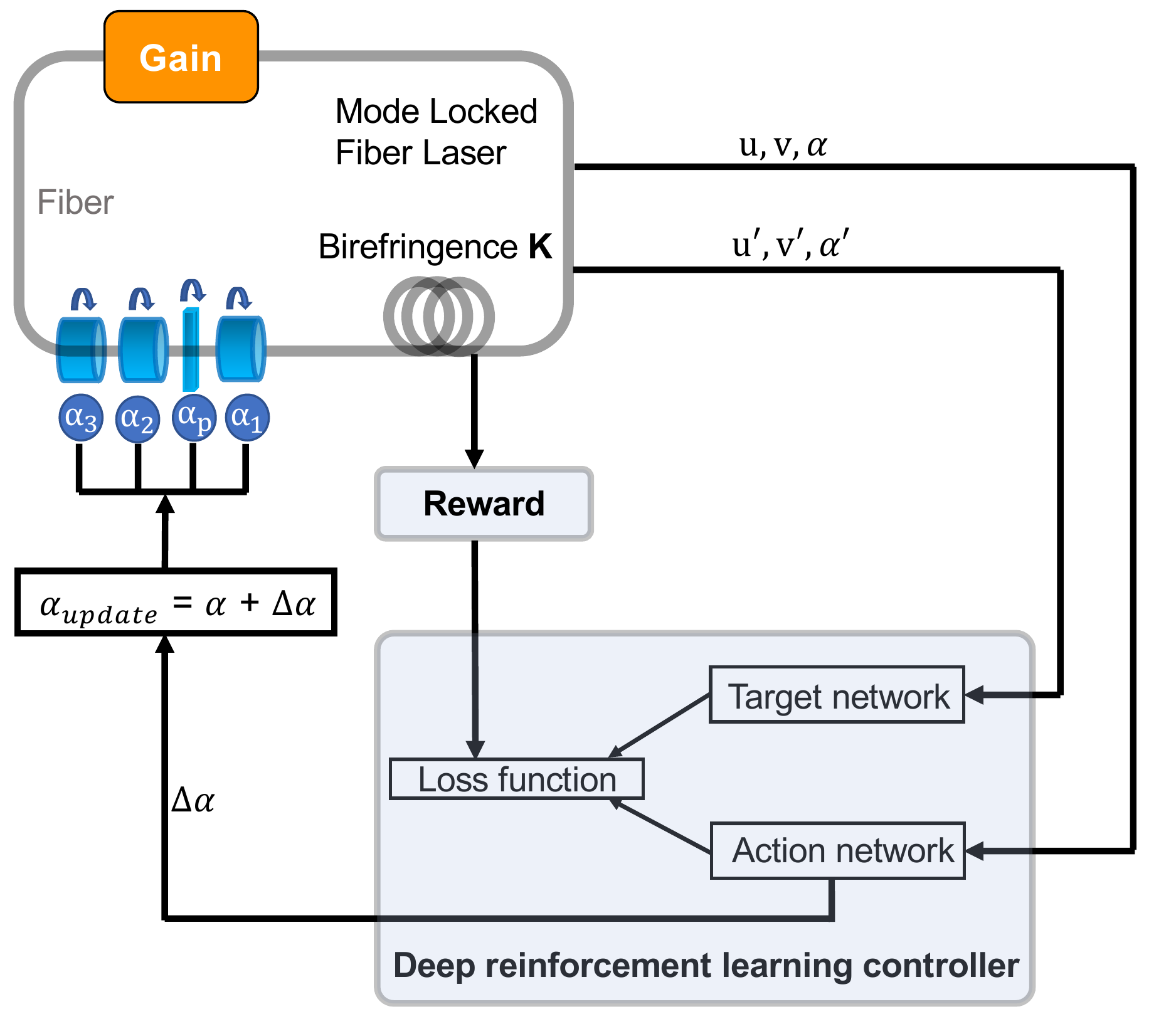}
 \vspace{-.2in}
  \caption{Schematic of the self-tuning fiber laser. The mode-locked fiber laser, including the laser cavity and optical components, is discussed in detail in the Methods Section \ref{sec:laser}. The deep reinforcement learning controller is discussed in the Methods Section \ref{sec:DRLC}.}
\label{fig:intro}
\end{figure}

In context of mode-locked lasers, the RL agent is given access to the components of the laser typically used for generating stable operation:  the waveplates and polarizer (See Fig.~\ref{fig:intro}).  The RL agent then explores the ways to maximize the policy information, which is centered around stable mode-locking of the laser cavity.  Specifically, the highest-energy mode-locked pulse is typically sought in the high-dimensional space generated by the waveplates and polarizer.  We show that the RL agent can learn to stabilize a mode-locked laser in a robust manner.  More than that, it can learn pathways to circumvent regions in parameter space where bi-stabilities exist.  Indeed, the delayed reward structure of the RL agent allows the system to learn how to maneuver around bi-stabilities in order to achieve optimal mode-locking performance.  Such a trajectory cannot be discovered with the variety of ML methods used so far on laser systems~\cite{brunton2013extremum,brunton2014self,fu2014classification,kutz2015intelligent,andral2015fiber,woodward2016towards,andral2016toward,baumeister2018deep}.  
The RL framework is especially valuable for systems where enough self-exploration can be promoted in order to sample the entirety of parameter space.  This can be done with mode-locked lasers so that the RL architecture provides a clear pathway towards technological implementation and more robust turn-key technologies.  

\section*{Results}

\begin{figure}[t]
 \centering
 \includegraphics[width=1.05\linewidth]{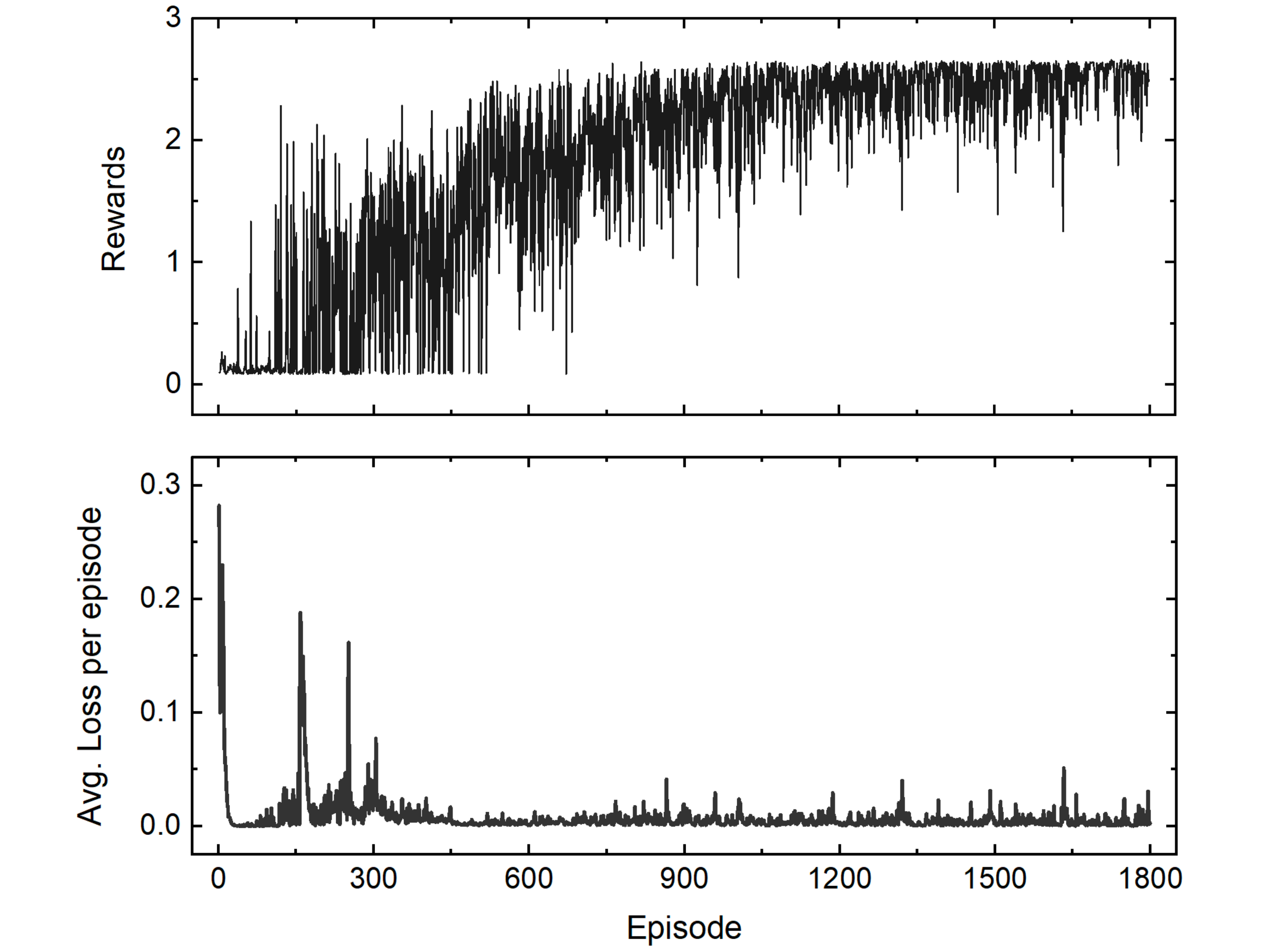}
 \put(-265,190){(a)}
 \put(-265,95){(b)}
 \vspace{-.2in}
 \caption{The variation of the rewards and loss function during training shows that the deep reinforcement learning controller adapts to improved policies as training proceeds.}
 \label{fig:single_1init_train}
\end{figure}

\begin{figure*}[t]
\vspace{-.2in}
\includegraphics[width=\linewidth]{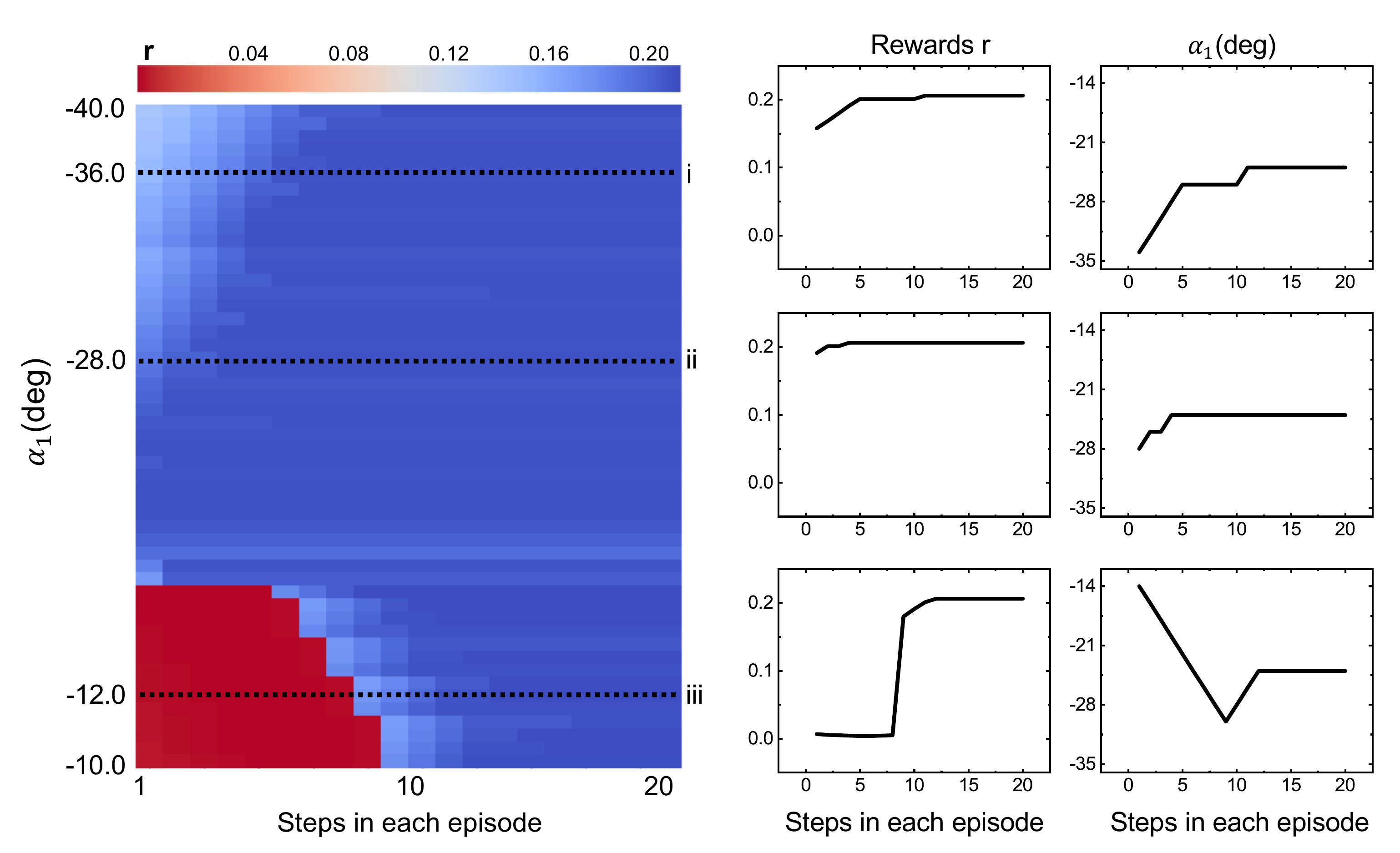}
\put(-255,283){(i)}
\put(-255,192){(ii)}
\put(-255,100){(iii)}
\vspace{-.2in}
\caption{The deep reinforcement learning controller effectively drives the laser dynamics to mode-locked solutions with $\alpha_1$ starting from $[-40^{\circ}, -10^{\circ}]$. Left panel demonstrates the change of rewards in each episode starting with different initial values of $\alpha_1$. The deep reinforcement learning controller adaptively selects actions to continue with the current waveplate orientation, or increase/decrease $\alpha_1$ by $2^{\circ}$. Control results for experiments starting with $\alpha_1=-36^{\circ}$, $-28^{\circ}$, and $-12^{\circ}$ are shown in detail in figures (i)-(iii). Note that the intra-cavity electric fields $u$ and $v$ start as hyperbolic secant pulses in each experiment.}
\label{fig:single_allinit}
\end{figure*}

We demonstrate the efficacy of deep reinforcement learning control on mode-locked fiber lasers in Fig.~\ref{fig:intro}. We first demonstrate the deep RL strategy for a single-input control ($\alpha_1$). The  deep RL controller is then applied in a multi-input control to find the optimal orientation of the waveplates ($\alpha_1, \alpha_2, \alpha_3$) and polarizer ($\alpha_p$). Finally, the controller is generalized to find optimal solutions with varying values of the fiber birefringence, which is an unmeasured latent variable that dictates the performance of the laser cavity.  RL is shown to be a robust and stable way to enact control.   The loss function, or optimization objective, is detailed in the Methods section.  Previous work~\cite{brunton2013extremum,brunton2014self} has found the loss function to be well modeled by the cavity energy divided by the kurtosis (fourth-moment) of the spectrum.

\subsection*{Single-input control for fixed birefringence}

Figure \ref{fig:single_1init_train} shows the variation of rewards and loss function during training process of the deep RL controller for a
single-input, single-output (SISO) case.  The quarter-waveplate angle $\alpha_1$ is the control variable, which can be varied in $2^{\circ}$ steps. The search starts with an initial value of $\alpha_1=15^{\circ}$, and all other angles are held fixed at pre-determined, locally
maximizing values. The deep reinforcement learning agent takes an action from a state using epsilon-greedy policy and the exploration rate $\epsilon$ exponentially decays during training. We observe an increase of the total reward over a complete episode as the training proceeds, as shown in Fig. \ref{fig:single_1init_train} (a). Note that the deep RL agent adapts to a new policy when the loss rises as shown in Fig. \ref{fig:single_1init_train} (b). 

\begin{figure}[t]
 \centering
 \vspace{-.1in}
 \includegraphics[width=\linewidth]{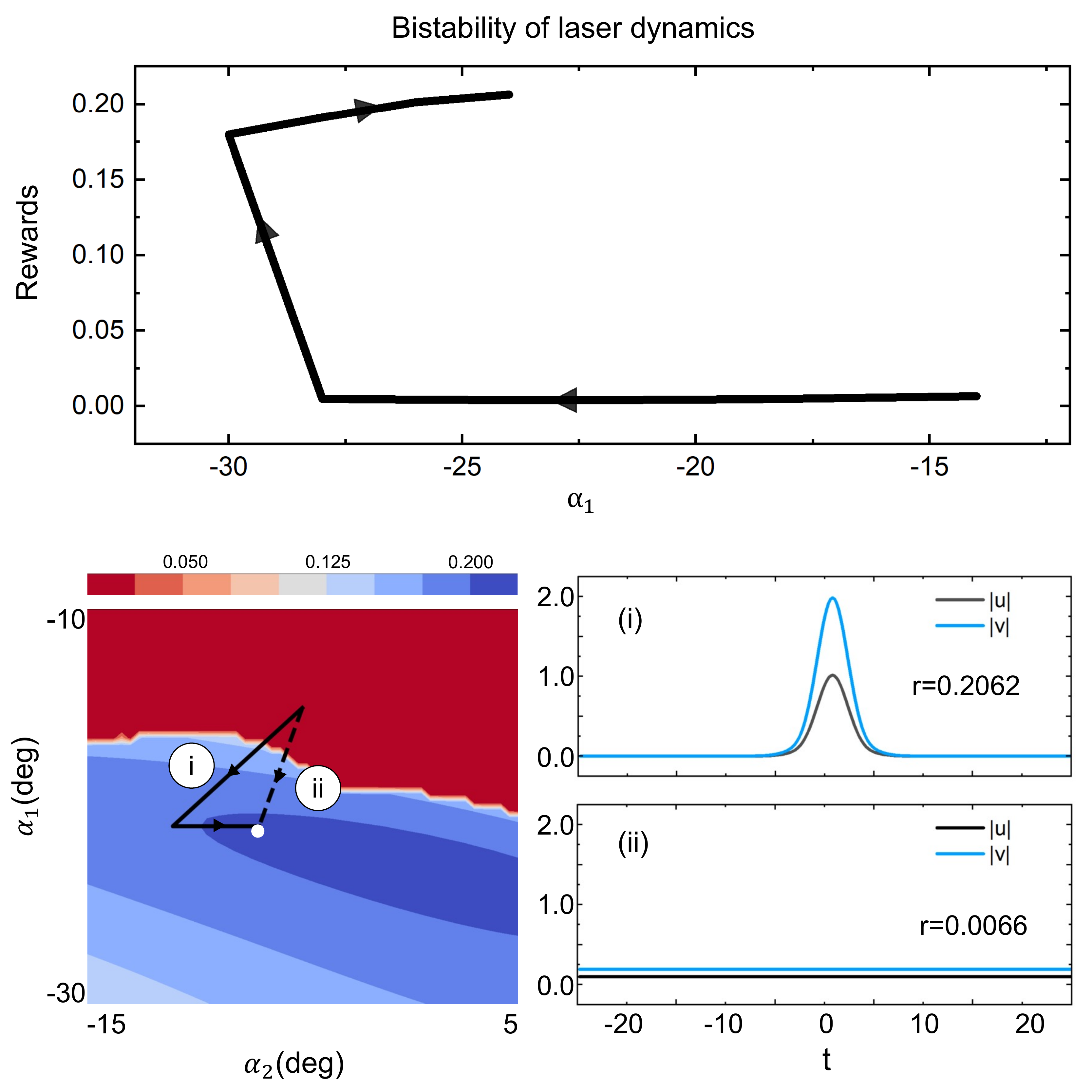}
 \put(-247,240){(a)}
 \put(-247,120){(b)}
 \vspace{-.1in}
 \caption{(a) The deep reinforcement learning controller for a single-input, single-output (SISO) case. The reward $r$ rises from the initial value as the controller drives the intra-cavity dynamics to mode-locking and we observe hysteresis in the corresponding change of $\alpha_1$ while the reward $r$ is consistently increasing. (b) The deep reinforcement learning control for two controllers $\alpha_1$ and $\alpha_2$. Start with $\alpha_1=-15^{\circ}$ and $\alpha_2=-3^{\circ}$, the laser dynamics successfully arrives at the mode-locked solution with $r=0.2062$ following the path (i) selected by the deep reinforcement learning controller, whereas we observe the plane wave solution ($r=0.0066$) following path (ii) as comparison.}
 \label{fig:bistability}
\end{figure}

\begin{figure}[t]
 \centering
 \vspace*{-.4in}
 \includegraphics[width=0.8\linewidth]{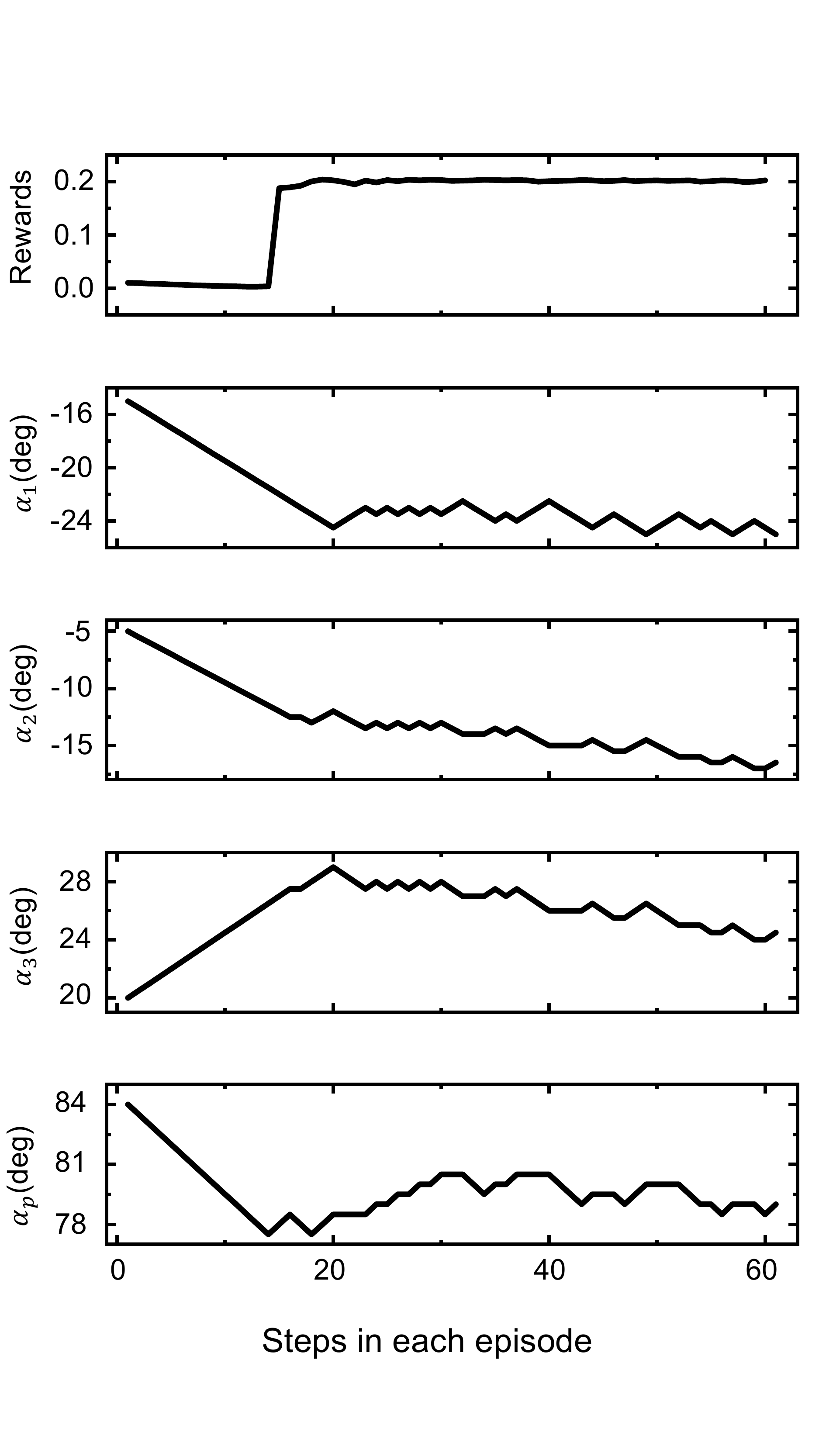}
 \vspace*{-.3in}
  \caption{The deep reinforcement learning controller for the multiple-input, single-output (MISO) case where we are controlling all four waveplate orientations simultaneously ($K=0$). The experiment starts with hyperbolic secant pulses $u$ and $v$ in cavity, which are promptly attenuated to constant waveforms with initial values of $\alpha_1=15^{\circ}, \alpha_2=-5^{\circ}, \alpha_3=20^{\circ}$, and $\alpha_p=84^{\circ}$. The four controllers $\alpha_1$, $\alpha_2$, $\alpha_3$, and $\alpha_4$ either hold on to the current orientations or increase/decrease by $0.5^{\circ}$ in each step.}
 \label{fig:four_1init}
\end{figure}

Extending the initial values of $\alpha_1$ from $15^{\circ}$ during training enables us to train a model that drives the laser dynamics to mode-locking with different initial values of $\alpha_1$, as shown in Fig. \ref{fig:single_allinit}. With fixed birefringence parameter $K$, the deep reinforcement learning controller correctly interprets and extracts features from the input states, and takes the action to efficiently drive the laser dynamics to mode-locked solutions with $\alpha_1$ starting from $[-40^{\circ}, -10^{\circ}]$. For example, with the initial value of $\alpha_1=-12^{\circ}$, the reward of each step continues increasing from the initial value as the deep RL controller drives the intra-cavity dynamics to mode-locking, as shown in Fig. \ref{fig:single_allinit} (iii).  
Interestingly, we observe hysteresis in the corresponding change of $\alpha_1$ while the reward is consistently increasing, as shown in Fig. \ref{fig:bistability} (a). The deep reinforcement learning controller correctly identifies the bi-stability of the intra-cavity dynamics and arrives eventually at the globally maximizing solution in this case.  No other ML architecture to date has been able to identify bistability.   RL achieves this due to its deferred reward structure which allows it to plan a path around the instability.  Figure \ref{fig:bistability} (b) describes the system bi-stability with two controllers $\alpha_1$ and $\alpha_2$. Our deep RL agent successfully discovers the path to drive the laser dynamics to mode-locking, marked as path (i) in Fig. \ref{fig:bistability} (b). The corresponding final states of the electric fields $u$ and $v$ are also pictured in Fig. \ref{fig:bistability} (b), with a high reward $r=0.2062$. We compare this path selected by our deep RL agent to the direct connection of the start and end points, which is marked as path (ii) in Fig. \ref{fig:bistability} (b). The corresponding final states of the electric fields $u$ and $v$ are also pictured, of which we observe constant waveforms (plane waves) with $r = 0.0066$. Note that in this work we only extend the initial values of $\alpha_1$ to be in the range of $[-40^{\circ}, -10^{\circ}]$ during training, but the deep RL agent has the ability to be generalized and further expanded to a larger range of initial values of $\alpha_1$.

\subsection*{Multi-input control for fixed birefringence}
Figure \ref{fig:four_1init} shows the deep reinforcement learning controller for
the multiple-input, single-output (MISO) case where we
control all four waveplate orientations simultaneously. This multi-input control is more complicated than the single-input case as the number of possible actions is significantly larger.  Thus transfer learning is leveraged to prevent the model from diverging at the early stage of training. The number of controllers is gradually increased until the desired performance is achieved. The search starts with initial values of $\alpha_1=15^{\circ}, \alpha_2=-5^{\circ}, \alpha_3=20^{\circ}$, and $\alpha_p=84^{\circ}$. With fixed fiber birefringence, the deep reinforcement learning controller takes the correct actions to drive the laser dynamics from constant waveforms (plane waves) to the mode-locked solutions, as shown in Fig. \ref{fig:four_1init}. After the mode-locked state is achieved, the deep reinforcement learning agent continues searching through the action space with the reward oscillating near the optimal performance. Because the waveplate orientations are varied simultaneously, the large action space results in a slow search for the deep RL agent, and we find it difficult to eliminate such oscillations. Other adaptive controllers, for example, extremum seeking control~\cite{brunton2013extremum}, can be combined to stabilize and attenuate such oscillations for better performance.


\subsection*{Multi-input control for varying birefringence}

\begin{figure}[t]
 \centering
 \includegraphics[width=\linewidth]{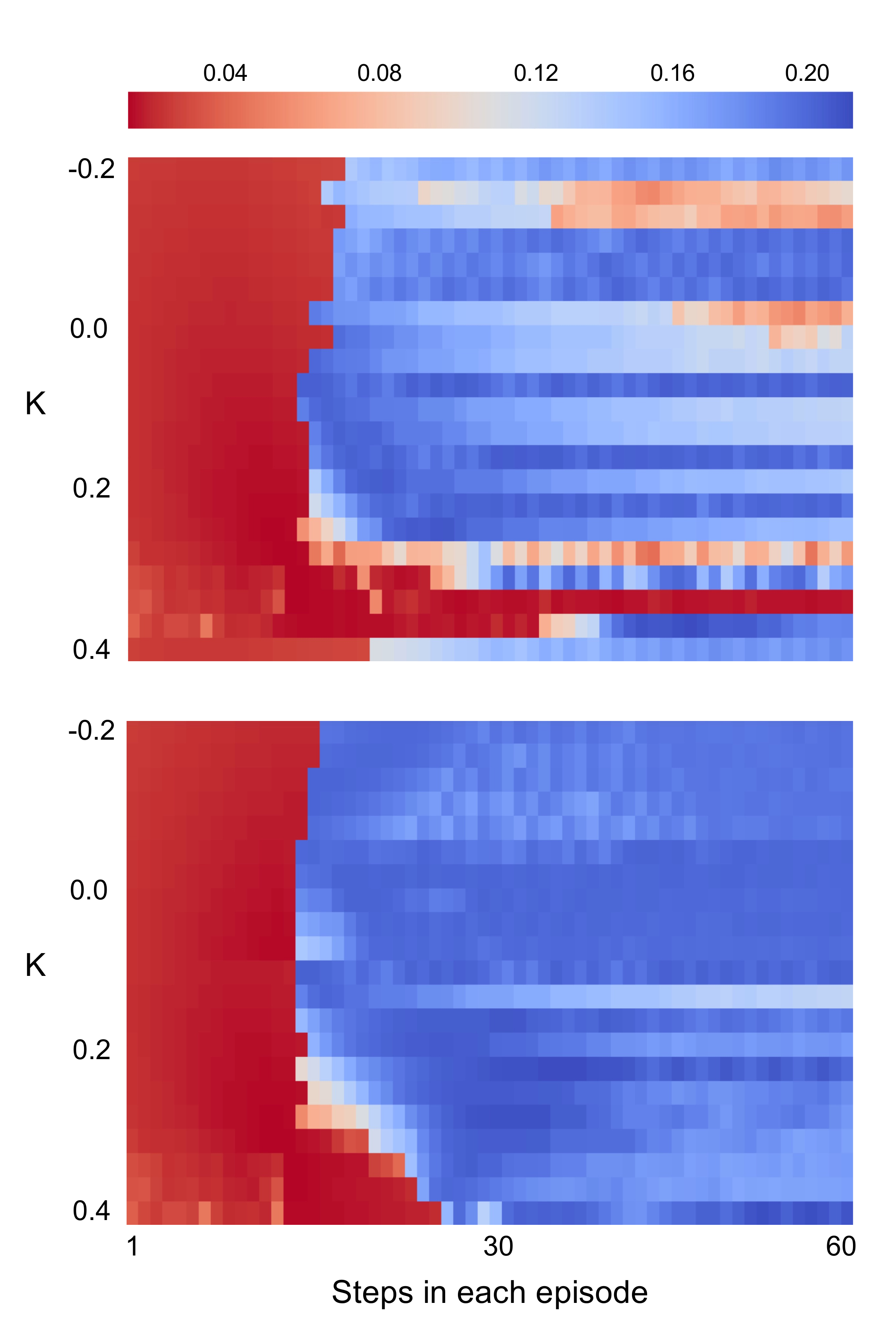}
  \put(-250,323){(a)}
 \put(-250,165){(b)}
 \vspace{-.2in}
 \caption{(a) Neural network parameters of the deep RL controller trained with birefringence $K=0.1$ can be generalized to environments with different values of $K$. (b) With transfer learning, neural network parameters of the deep RL controller can be rapidly fine-tuned with a small amount of newly collected experiences and updated to control effectively in the new environments with different values of $K$.}
 \label{fig:generalK}
\end{figure}

We find the neural network parameters of the deep RL controller trained with birefringence $K=0.1$ can be generalized to control  the mode-locking dynamics at different values of $K$. Except for a few cases, the deep reinforcement learning controller successfully drives the laser dynamics to mode-locking with different values of birefringence $K$ from $-0.2$ to $0.4$, as shown in Fig. \ref{fig:generalK} (a). Note that in such generalized cases, the deep RL controller takes more steps on average to drive the laser dynamics to mode-locking, and in some cases the mode-locked solutions achieved are not tightly-confined as shown previously. Moreover, among a few cases, the deep reinforcement learning controller successfully achieves mode-locking but gradually wanders afterwards. 

One feasible solution to improve the control performance in such cases is to retrain the model completely for different values of birefringence $K$. However, we find the neural network parameters of our deep RL agent adapt well and quickly with transfer learning~\cite{pan2009survey} to different values of the birefringence $K$. In other words, there is no need to retrain the model completely with varying $K$, but instead we can slightly increase the exploration rate of the current model and collect more experiences by interacting with the new environment of changed birefringence $K$. These newly collected experiences enable the model to quickly update and adapt to the new environment.  Thus it provides the possibility of building an online model for what is typically a stochastic and slowly-varying birefringence. In our experiments, training the neural network parameters with the birefringence $K=0.1$ takes at least 2000 episodes for the deep RL controller to converge to the optimal policy, whereas the transfer learning takes only 300 episodes on average to adapt to environments with varying birefringence $K$. With fined-tuned parameters for $K\in\{-0.2,0,0.2,0.4 \}$, we improve the performance of the deep reinforcement learning controller to drive the laser dynamics to mode-locking with other values of the birefringence $K$ ranging from $-0.2$ to $0.4$, as shown in Fig. \ref{fig:generalK} (b). Note only four sets of neural network parameters are used to generalize the controller to a range of $K$ values.



\begin{figure}[t]
 \centering
\includegraphics[width=\linewidth,trim=2.5cm 0 2.7cm 0, clip=true]{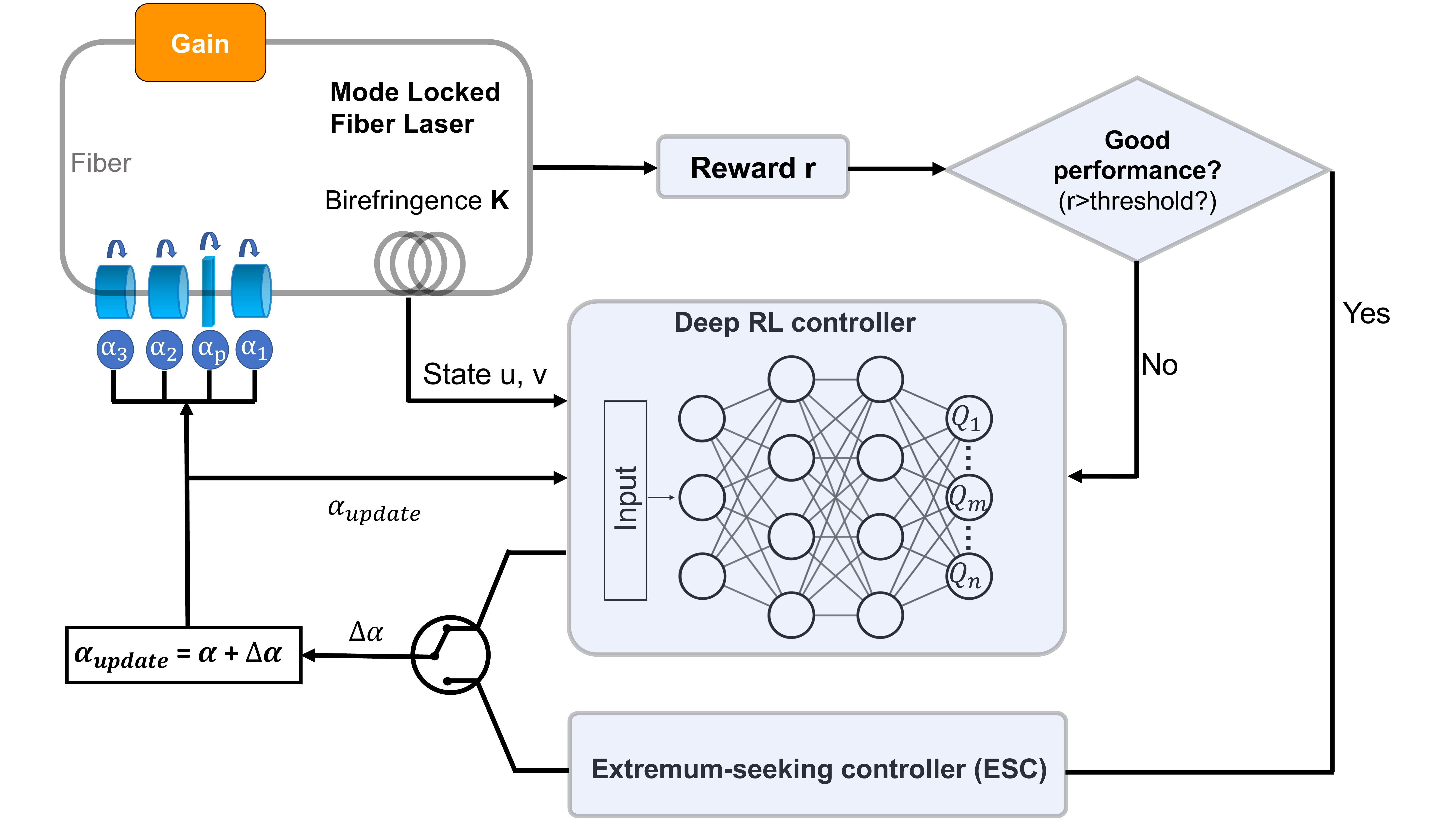}
\vspace{-.3in}
 \caption{Combining the deep reinforcement learning controller with the extremum-seeking controller stabilizes the intra-cavity dynamics and achieves mode-locking with varying birefringence $K$.}
 \label{fig:vary}
\end{figure}

As previously noted, we find in some cases that the deep RL controller successfully achieves mode-locking, but gradually walks apart from the desired solution. In such cases we can rely on extremum seeking controller \cite{brunton2013extremum} or other adaptive controllers for better performance. Extremum-seeking control, for example, is a form of perturb-and-observe control that estimates the gradient of an objective function by injecting an additional sinusoidal signal as input. The signal converges more rapidly when the objective function has a large gradient. Extremum-seeking control can lock the system to the local maximum and reacts rapidly to moderate changes of intra-cavity dynamics~\cite{krstic2000stability}. However, it relies on initial conditions of the parameters and state of the system since it only finds local maxima. Moreover, extremum-seeking control cannot recover in cases when the system is knocked far from the desired local maximum with drastic
perturbations. Therefore, a combination of the deep RL agent with the extremum-seeking controller is a viable integrative strategy since it can evade the poor local maximum, and stabilize the intra-cavity dynamics around the mode-locked solution.  To implement this integrated strategy, the deep RL controller is first executed in order to find a good mode-locked solution in a rapid manner.  Indeed, deep RL can find near optimal stable mode-locking with several steps of propagation.  The extremum-seeking controller is then turned on to stabilize the system. The schematic is shown in Fig. \ref{fig:vary}.

\section*{Discussion}

Deep reinforcement learning is a learning paradigm that integrates  the power of reinforcement learning and deep neural networks.  It is an ideal ML paradigm for complex dynamical systems where the learning agent is allowed to explore the system and for which trajectory planning is critical for success:  both aspects typically manifest in optical systems.
Here, we demonstrate a fast, reliable self-tuning controller for the passive mode-locked fiber laser with deep RL. The controller varies all four waveplate orientations simultaneously to achieve a tightly-confined, high-energy mode-locked state. Interestingly, the control paths selected by the deep reinforcement learning controller reflect the bi-stability of the laser dynamics, and demonstrate the efficacy of the deep learning control to correctly sense the state of the environment in a bistable system. Although no new optical physics is demonstrated in this work, we have provided a principled control strategy to escape from the poor local maxima by interacting with the environments, which is important for building controllers in systems with bi-stabiltiy. Moreover, the deep RL controller architecture provided here can be easily integrated and generalized to experimental environments and other optical systems, including for instance, managing instabilities from dispersion management~\cite{bronski1996modulational}, controlling pulse compression~\cite{li2013high}, and/or circumventing Q-switching instabilities~\cite{proctor2006theory}.  Importantly, given a well-defined reward criteria and state-space, the deep RL architecture generates experiences with its environment in order to train the deep reinforcement learning controller.  

The deep RL framework demonstrated here can be combined and integrated with other control paradigms, for example, the extremum-seeking controller, for a better control performance. Since the deep RL controller continues searching the entire space even with a good mode-locked solution already found, it is difficult to eliminate the oscillation, and in some cases the controller eventually walks apart from the mode-locked solutions. An integration with the extremum-seeking controller stabilizes the control performance around the optimal solution discovered, even with slowly varying birefringence. With drastic perturbations to the birefringence, our deep reinforcement learning controller can be promptly fine-tuned to adapt to the new environments using transfer learning. Once a new mode-locked solution is found by the deep RL controller, the extremum-seeking controller is turned on instead to stabilize the system.  This hybrid approach marries the ability of deep RL to search globally in a large control space with the increased stability provided by extremum-seeking control via local optimization.

\section*{Methods}

\subsection{Reinforcement learning}
As noted earlier, RL is a branch of machine learning that uses a goal-oriented algorithm that learns from interactions with its environment.   Using a trial-and-error search, an agent learns to sense the state of its environment and take actions accordingly to achieve optimal immediate or delayed rewards. Specifically, the RL agent arrives at different states by performing actions, with the selected actions leading to positive or negative rewards for learning. The agent's behaviors are defined by policies of the reinforcement learning algorithms in the environments, and we target at the optimal policies for reinforcement learners to maximize total rewards across an episode (or trajectories generated by tuning the optical system).  

\subsubsection*{$Q$-learning}
We leverage deep $Q$-learning, specifically the deep off-policy temporal difference control algorithm \cite{Q_learning, DQ_learning}, which approximates the current estimations based on the previously learned estimations. In reinforcement learning algorithms, the state-action value function, or $Q$-function, is defined as the expected discounted return of rewards starting from the state $s$ with the action $a$ according to policy $\pi$. The $Q$-function specifies the agent's performance taking a particular action and transit from the current state to the next with the policy we choose. During training the reinforcement learning agent learns and converges to the optimal policy that maximizes the total reward across an episode. 

$Q$-learning~\cite{WatkinsThesis} is a particular approach to learn optimal actions in such sequential decision problems and has been recognized as a form of temporal difference learning~\cite{Sutton1988}.
Suppose we take action $a$ in the current state $s$ and arrive at state $s'$, $Q$-learning obtains 
\begin{equation}
Q(s,a) = r(s,a)+\gamma \max_{a'}Q(s', a'),
\end{equation}
where $r(s,a)$ is the reward collected performing action $a$ to move from state $s$ to $s'$, $\gamma\in[0,1]$ is the discount factor that controls the contribution of the rewards collected in the future to the total reward after the episode is finished. 
During an episode, the agent proceeds by either choosing the action with the highest $Q$ value (exploitation), or selecting randomly an action to explore other possible states which may return higher delayed rewards (exploration). The agent moves forward to the next state $s'$ with the selected action $a$ and collects the associated reward $r$. 
%
$Q$-learning updates the current $Q$ value of the experienced state-action pair with the collected reward after transitioning to $s'$ and the possible future rewards taking the optimal action thereafter:
 \begin{equation}
 \small   Q^{new}(s,a) = Q(s,a)+\alpha (r+\gamma \max_{a'}Q(s', a')-Q(s,a)),
 \label{eq:Q_learning}
 \end{equation}
where $\alpha\in[0,1]$ is the learning rate. Note that the difference between the actual reward $r+\gamma \max Q(s', a')$ and the expected reward $Q(s,a)$ is taken to update the value of $Q(s,a)$. The parameter $\alpha$ is important for convergence since it determines to what extent the current $Q$ function is updated by the newly explored information. 
The $Q$ function is arbitrarily initialized and updated following Eq.~(\ref{eq:Q_learning}) until the $Q$-learning algorithm has converged. 

\subsubsection*{Deep Q neural networks (DQN)}
\label{sec:DQN}
In discrete environments represented by a finite number of possible states and actions, 
we often search through all possible 
state-action pairs exhaustively to find the optimal $Q(s,a)$ values and the associated policy. 
However, this is computationally expensive and becomes infeasible with more than a small number of state-actions pairs. In continuous environments, it is impossible to list and search through each state with different actions. 
In contrast, deep $Q$ learning \cite{DQ_learning} allows one to approximate the tabular $Q$ function $Q(s,a)$ as a parameterized function $Q(s,a;\theta)$.
Considering that neural networks can provide good approximations to possibly very complex functions, we utilize here deep neural networks as the estimator of the $Q$ value function. 
In particular, $Q(s,a;\theta)$ is modeled as a multi-layered neural network with parameters $\theta$ that takes a given state $s$ as input and yields a vector of values $Q(s,\cdot;\theta)$, each associated with a particular action $a$.

Following the $Q$-learning updating rule defined in Eq. \ref{eq:Q_learning}, we refer to $r+\gamma \max Q(s',a';\theta)$ as the target value, $Q(s,a;\theta)$ as the predicted value, and the difference between the target and prediction is minimized when the current policy converges to the optimum. In deep $Q$ learning, we can define analogously the loss function as the squared difference between the target and predicted value:
\begin{equation}
L = (r+\gamma \max_{a'}Q(s',a';\theta)-Q(s,a;\theta))^2.
\label{eq:DDQN} 
\end{equation}
The loss is minimized by learning updates to the deep neural network parameters $\theta$ that converge to the optimal policy. In summary, we use neural networks for the approximation of the $Q$ function in deep $Q$ learning, and converge to the optimal policy by minimizing the loss. 

\begin{figure}[t]
 \centering
 \includegraphics[width=\linewidth]{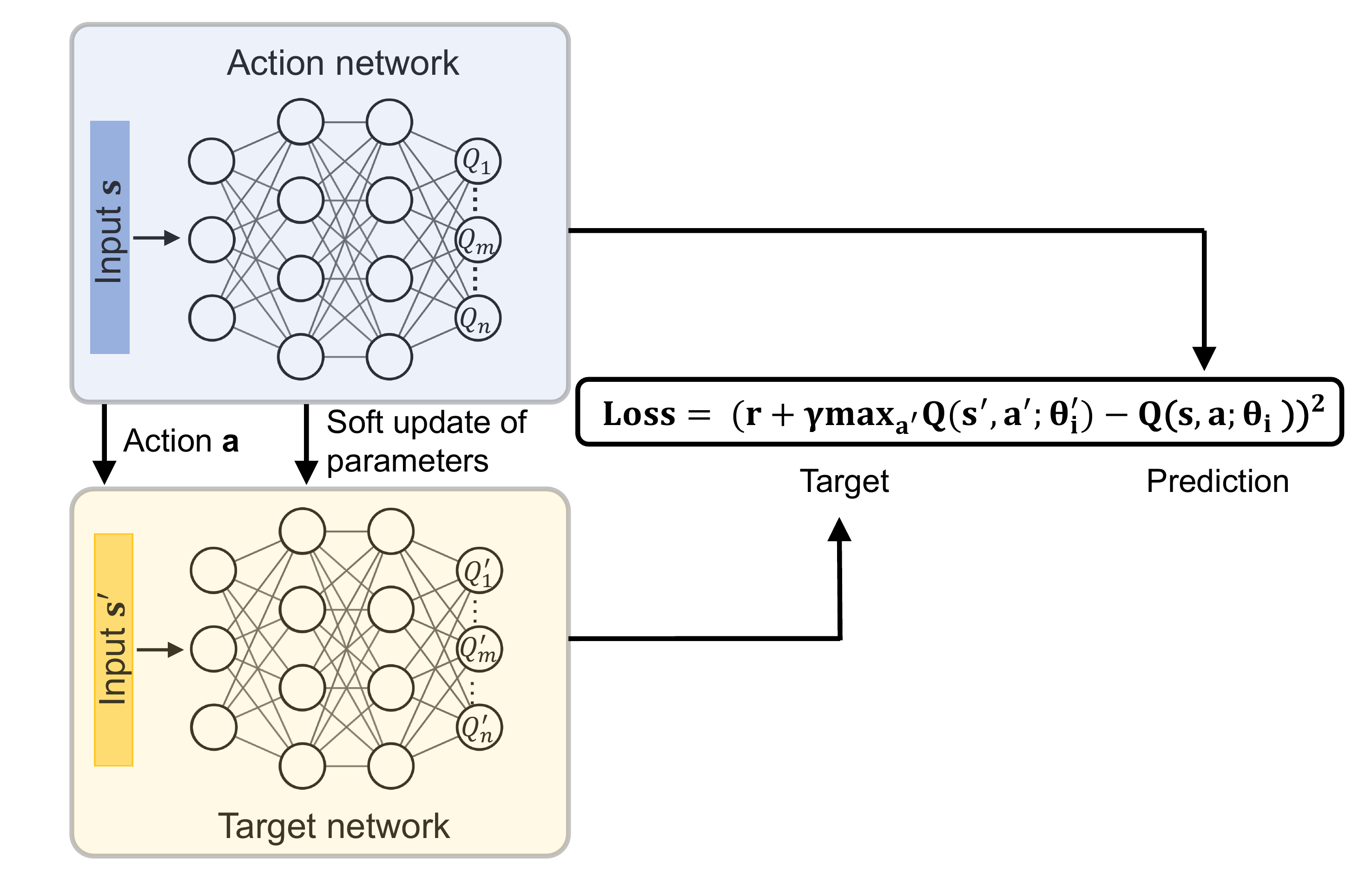}
 \vspace{-.3in}
 \caption{The architecture of the double deep Q neural network. A target network is included to stabilize the training. More details are discussed in Section \ref{sec:DQN}.}
 \label{fig:DDQN}
\end{figure}

In particular, we employ as deep reinforcement learning agent an adaptation of the double deep $Q$ neural network (DDQN) \cite{DoubleQ_learning, DDQ_learning} to the self-tuning laser control problem. The architecture of DDQNs is shown in Fig. \ref{fig:DDQN}, 
where the inputs fed into the action network describe the current state that the deep RL agent is in, and the output of the action network is the approximated $Q$ function, specifically the $Q$ values for all possible actions given the current state. 
Following the loss function defined in Eq.~(\ref{eq:DDQN}), we would observe strong divergence during training since the same neural network with parameters $\theta$ calculates both the
predicted value and target value. To diminish the divergence, two separate networks are employed, one for selecting an action and the other for evaluating the selected action.
Specifically, the target network with parameters $\theta '$ is used to calculate the target value, while the action network with parameters $\theta$ yields the predicted Q values associated with each action. The new loss function is defined as:
\begin{equation}
L =  (r+\gamma \max_{a'}Q(s', a'; \theta ')-Q(s,a;\theta))^2,
\end{equation}
where $\theta '$ and $\theta$ stand for the different set of parameters of the target network and the action network, respectively. The parameters of the target network are periodically frozen for several episode during training before being updated by copying the parameters from the action network, or partially updated with parameters from the action network in each episode to stabilize the training.

\begin{figure*}[t]
\includegraphics[width=\linewidth]{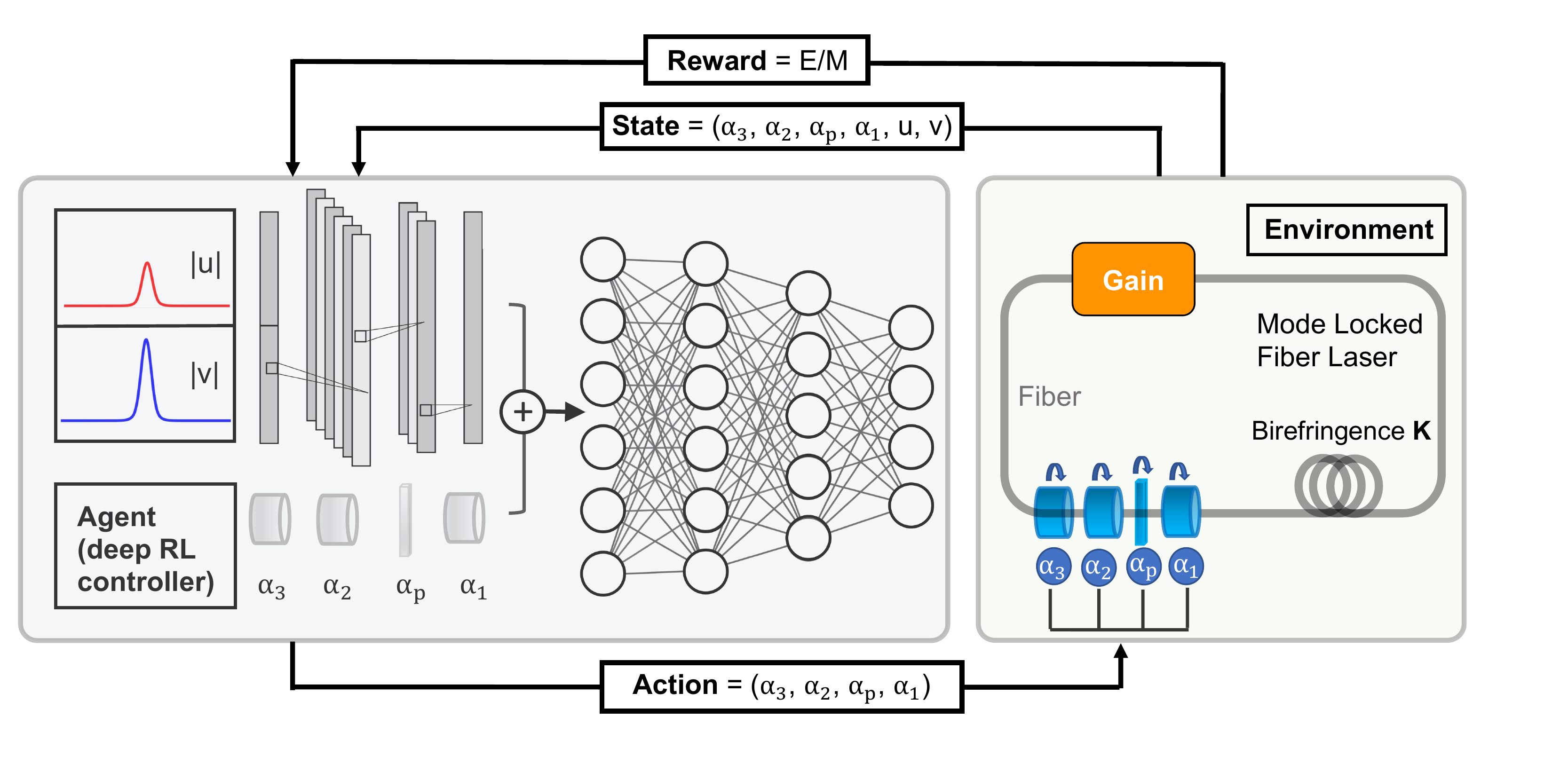}
\vspace{-.3in}
\caption{Schematic of the self-tuning mode-locked laser with deep reinforcement learning control. The input to the deep RL controller describes the current state that the controller is in, and is defined as the concatenated electric fields $u$, $v$ and waveplate orientations $\alpha_1$, $\alpha_2$, $\alpha_3$, $\alpha_p$. The control inputs to the laser cavity are then updated by the selected action of the deep RL controller, which result in changes of the laser cavity dynamics and returns new electric fields $u$, $v$, and reward $r$ as defined in Eq. \ref{eq:reward} to the deep RL agent. Given the updated inputs and the associated reward $r$, the deep RL controller adjusts its strategy accordingly to select the next action and optimize the control inputs to the laser cavity.}
\label{fig:DRLML}
\end{figure*}

To stabilize the training and reduce the
overfitting caused by correlation between the 
deep RL agent's experiences, we train the DDQN with an experience replay buffer \cite{lin1993reinforcement}, which is usually defined as a queue that saves a fixed number of the recent experiences. The experience $<s,a,r,s'>$ of the deep RL agent is defined as the concatenation of the current state $s$, the action selected $a$, the next state $s'$ after performing the action, and the associated reward $r$ received in this transition. During training, rather than directly train with the newest experiences collected, we sample a random batch of the experiences $<s,a,r,s'>$ from the replay 
buffer and feed the sampled batch to the neural network for parameter updates. The deep RL agent benefits from the replay buffers by learning from an enlarged range of random and less correlated experiences. 
 
\subsection{Mode-locked fiber laser model}
\label{sec:laser}

Our model of the laser cavity is a well-established computational model which treats the cavity in a component by component manner by separately applying the nonlinear optical propagation to the laser dynamics with discrete waveplates and polarizer in each round trip.  This model produces a rich set of dynamics that we wish to control~\cite{spaulding2002nonlinear}.
We model the propagation of intra-cavity fields with the coupled nonlinear Schr\"odinger equation with modifications to account for the bandwidth limited gain and cavity losses~\cite{ding2011generalized, ding2011operating, komarov2005multistability}:
\begin{subequations}
	\begin{align}
\footnotesize\hspace{-0.2cm} i\frac{\partial u}{\partial z}\hspace{-0.05cm}+\hspace{-0.05cm}\frac{D}{2}\frac{\partial^2 u}{\partial t^2}\hspace{-0.05cm}-\hspace{-0.05cm}Ku\hspace{-0.05cm}+\hspace{-0.05cm}(|u|^2\hspace{-0.05cm}+\hspace{-0.05cm}A|v|^2)u\hspace{-0.05cm}+\hspace{-0.05cm}Bv^2u^* &= iRu,\\
\footnotesize\hspace{-0.2cm} i\frac{\partial v}{\partial z}\hspace{-0.05cm}+\hspace{-0.05cm}\frac{D}{2}\frac{\partial^2 v}{\partial t^2}\hspace{-0.05cm}+\hspace{-0.05cm}Kv\hspace{-0.05cm}+\hspace{-0.05cm}(A|u|^2\hspace{-0.05cm}+\hspace{-0.05cm}|v|^2)v\hspace{-0.05cm}+\hspace{-0.05cm}Bu^2v^* &= iRv,
	\end{align}
\end{subequations}
where $u(z, t)$ and $v(z, t)$ are often referred to as the fast and slow components of the two intra-cavity electric field envelopes, which are orthogonally polarized. 
The propagation distance $z$ is non-dimensionalized by the cavity length, and the dimensionless time $t$ is normalized by the full width at half maximum of the pulse. $D$ is the average group velocity dispersion, $A$ and $B$, determined by physical properties of the laser fiber, are the nonlinear coupling parameters characterizing the cross-phase modulation and the four-wave mixing, respectively. In this work we consider a silica fiber with $A = 2/3$ and $B = 1/3$. The fiber birefringence, quantified by $K$, represents a major disturbance to the laser dynamics due to its sensitivity to thermal fluctuations. The dissipative term $R$, characterizing the bandwidth-limited gain and attenuation
arising from the Yb-doped amplification, is defined as 
\begin{equation}
R=\frac{2g_0(1+\tau\partial_t^2)}{1+(1/e_0)\int_{-\infty}^{\infty}(|u|^2+|v|^2)dt}-\Gamma,
\end{equation}
where $g_0$ is the dimensionless pumping strength, and $e_0$
is the dimensionless saturation energy of the gain medium. Losses caused by output coupling and fiber attenuation are characterized by the pump bandwidth $\tau$ and $\Gamma$.

The effect of the waveplates and polarizer during each round trip is modeled by the
discrete application of Jones matrices:
\begin{equation}
W_{\lambda/4}=\left[
\begin{matrix}
e^{-i\pi/4} & 0 \\
0 & e^{i\pi/4}
\end{matrix}
\right], \tag{7.1}
\end{equation}
\begin{equation}
W_{\lambda/2}=\left[
\begin{matrix}
-i & 0 \\
0 & i
\end{matrix}
\right], 
W_p=\left[
\begin{matrix}
1 & 0 \\
0 & 0
\end{matrix}
\right]. \tag{7.2}
\end{equation}
Note that $W_{\lambda/4}$ characterizes the effects of quarter-waveplates $\alpha_1$ and $\alpha_2$, $W_{\lambda/2}$ is
 for the half-waveplate $\alpha_3$, and $W_p$ is for the polarizer $\alpha_p$. An additional rotation matrix $R(\alpha)$ is necessary to account for the offset between the direction of the intra-cavity fast field and the principal axes of the waveplates and polarizer, and we define
\begin{equation}
J_j = R(\alpha_j)W_jR(-\alpha_j), \tag{8.1}
\end{equation}
\begin{equation}
R(\alpha_j)=\left[
 \begin{matrix}
   cos(\alpha_j) & -sin(\alpha_j) \\
   sin(\alpha_j) & cos(\alpha_j)
  \end{matrix}
  \right],\tag{8.2}
\end{equation}
where $\alpha_j$ $( j = 1, 2, 3, p)$ is a waveplate or polarizer angle. These rotation angles are easily manipulated via electronic control \cite{shen2012electronic}, and are considered as the control variables of the deep reinforcement learning agent for driving the laser dynamics to mode-locking in this work.

\subsection{Deep reinforcement learning control}
\label{sec:DRLC}
A schematic of the self-tuning mode-locked laser with deep reinforcement learning control is shown in Fig. \ref{fig:DRLML}, highlighted with a deep RL controller and a mode locked fiber laser cavity of passive nonlinear polarization
rotation (NPR). The mode-locking laser cavity, which is discussed in details in the previous section, is interpreted as the interactive environment in the reinforcement learning framework, and the waveplate angles $\alpha_1, \alpha_2, \alpha_3$ and polarizer angle $\alpha_p$ are considered as the controllable actions of the deep reinforcement learning agent. We take concatenated components of the electric fields $u$, $v$, and the current waveplate orientations $\alpha_1, \alpha_2, \alpha_3$ and $\alpha_p$ as the input to the deep reinforcement learning agent. The deep reinforcement learning controller is built with alternatively stacked convolutional layers and max pooling layers, followed by fully connected layers with leaky-ReLU as activation functions. The convolutional layers extract features from the input state by identifying the solitons inside the electric fields $u$ and $v$, and the max pooling layers detect existence of the solitons and reduce the input dimensionality before feeding into the fully connected layers. Note that we demonstrate in this work the efficacy of the deep RL architecture in a numerical simulation of the laser cavity, but it is possible to train the deep RL controller directly in an experiment, as the controller only relies on information that is readily available in experiments. 

The performance of the deep reinforcement learning controller is evaluated in terms of a reward $r$. In particular, we seek to steer the system to high-energy mode-locked states. However, the reward/cost landscape is very complex and exhibits many local optima. In addition, evaluating energy is not sufficient, as there are many chaotic solutions which have significantly higher energy than mode-locked states~\cite{brunton2013extremum}. To define the reward $r$, we consider including the fourth-moment (kurtosis) $M$ of the Fourier spectrum of the waveform, which is large for chaotic solutions but relatively small for the desired mode-locked states. To have a large reward $r$ only for tightly confined temporal wave packets with relatively large energy, we define~\cite{brunton2013extremum}
\begin{equation}
r = E/M.\tag{9}
\label{eq:reward}
\end{equation}
To penalize the ineffective actions more efficiently during training, we rescale the reward to be centered around zero, so that the desired actions result in positive rewards while the ineffective ones return negative rewards. We rescale it back as defined in Eq.~(\ref{eq:reward}) after training for consistency and interpretability.

Our deep RL agent uses an $\epsilon$-greedy policy to balance between exploration and exploitation, and parameters $\theta '$ of the target network are partially updated in each training step to improve stability. Note that the deep RL agent spans a large action space in the multiple-input single-output (MISO) case, when the three waveplates and polarizer orientations $\alpha_1$, $\alpha_2$, $\alpha_3$, and $\alpha_p$ are considered as controllers.  Thus we observe convergence difficulty in training the model directly with randomly initialized neural network parameters. To deal with this problem, we start training our deep RL agent with a single controller $\alpha_1$, and gradually increase the number of controllers by initializing with previously trained parameters for the current model  of increased number of controllers. Such parameter initialization strategy efficiently prevents the model from diverging, especially at the early stages of training.  

The deep RL agent takes as input the electric fields and waveplate orientations, and the output yields $Q$ values associated with all possible actions given the current state that the deep RL agent is in. Following the $\epsilon$-greedy policy, our RL agent either randomly selects an action for exploration, or greedily selects the action with the highest Q value for exploitation and moves to the next state accordingly. Specifically, parameters $\alpha_1$, $\alpha_2$, $\alpha_3$, and $\alpha_p$ are adjusted according to the action selected, and consequently we observe changes of electric fields $u$ and $v$ in the fiber laser cavity. The reward $r$ of this transition, as defined in Eq. \ref{eq:reward}, is taken of the new intra-cavity fields $u$ and $v$ after transition.  The procedure is repeated until the completion of the entire episode, and then a new episode is started with the same initial conditions of the electric fields and waveplate/polarizer orientations to collect more samples. Thus, the deep RL agent learns from different trials using exploration and exploitation, and eventually converges to the optimal policy that leads to the highest total reward across the entire episode. Note that after the training stage, the learned policy of the deep RL agent is evaluated by greedily selecting the action with the highest $Q$ value. 

~



\section*{Acknowledgements}
SLB acknowledges support from the Army Research Office (ARO W911NF-19-1-0045; Program Manager Matthew Munson).  SLB and EK acknowledge support from the Army Research Office (ARO W911NF-17-1-0306; Program Managers Matthew Munson and Samuel Stanton).  SLB, EK, and JNK acknowledge support from the UW Engineering Data Science Institute, NSF HDR award \#1934292.  
JNK acknowledges support from the Air Force
Office of Scientific Research (FA9550-17-1-0200).

\bibliographystyle{unsrt}
\bibliography{refs}

\end{document}